\documentclass[journal]{IEEEtran}
\IEEEoverridecommandlockouts
\usepackage[top=0.74in, bottom=0.7in, left=0.6in, right=0.6in]{geometry}

\usepackage{amsfonts}
\usepackage{amsmath}
\usepackage{amssymb}
\usepackage{algorithm}
\usepackage{algorithmicx}
\usepackage{algpseudocode}
\usepackage{dsfont}
\usepackage{authblk}
\usepackage{bbm}
\usepackage{graphicx}
\usepackage{epstopdf}
\usepackage{subfigure}
\usepackage{stfloats}
\usepackage{eqnarray}
\usepackage{makecell}
\usepackage{multirow}
\usepackage{enumerate}
\usepackage{booktabs}
\usepackage{cite}
\usepackage{balance}
\usepackage{color}
\usepackage{bm}
\usepackage{caption}
\usepackage{url}

\renewcommand{\IEEEQED}{\IEEEQEDopen}

\newcommand{\Tr}{\mbox{Tr}}

\newcommand{\REQ}{\mbox{\tiny REQ}}
\newcommand{\TOT}{\mbox{\tiny TOT}}
\newcommand{\SEC}{\mbox{\tiny SEC}}

\newcommand{\OPT}{\mbox{\tiny OPT}}
\newcommand{\SINR}{\mbox{SINR}}

\renewcommand{\H}{\mbox{\tiny H}}
\newcommand{\F}{\mbox{\tiny F}}
\newcommand{\Rank}{\mbox{Rank}}
\newcommand{\C}{\mbox{\tiny CIR}}
\newcommand{\Q}{\bm Q}
\newcommand{\W}{\bm W}
\newcommand{\sumn}{\sum\limits_{n=1}^N}
\newcommand{\sumnk}{\sum\limits_{k=1, k \neq n}^N}
\newcommand{\sumk}{\sum\limits_{k=1}^N}

\renewcommand{\(}{\left(}
\renewcommand{\)}{\right)}
\renewcommand{\[}{\left[}
\renewcommand{\]}{\right]}

\newtheorem{lemma}{\textbf{Lemma}}

\newtheorem{remark}{\textbf{Remark}}
\newtheorem{proposition}{\textbf{Proposition}}

\begin{document}
\title{\LARGE Robust Secrecy Energy Efficient Beamforming in MISOME-SWIPT Systems With Proportional Fairness}

\author[$\ddag$]{\mbox{Yanjie Dong},~\IEEEmembership{Student Member, IEEE}}
\author[$\dag$]{\mbox{Md. Jahangir Hossain},~\IEEEmembership{Senior Member, IEEE}}
\author[$\dag$]{\mbox{Julian Cheng,~\IEEEmembership{Senior Member, IEEE}}}
\author[$\ddag$]{\mbox{Victor C. M. Leung},~\IEEEmembership{Fellow, IEEE}\vspace{-0.2 cm}}
\affil[$\ddag$]{Department of Electrical and Computer Engineering\\
The University of British Columbia\\
Vancouver\\
\mbox{BC}\\
Canada}
\affil[$\dag$]{School of Engineering\\
The University of British Columbia\\
Kelowna\\
BC\\
Canada
\authorcr Emails: \{ydong16, vleung\}@ece.ubc.ca\\
\{julian.cheng, jahangir.hossain\}@ubc.ca\vspace{-1 cm}}

\maketitle
\pagestyle{empty}
\thispagestyle{empty}
\begin{abstract}
The joint design of beamforming vector and artificial noise covariance matrix is investigated for multiple-input-single-output-multiple-eavesdropper simultaneous wireless information and power transferring \mbox{(MISOME-SWIPT)} systems.
A secrecy energy efficiency (SEE) maximization problem is formulated in the \mbox{MISOME-SWIPT} system with imperfect channel state information and proportional secrecy rate constraints.
Since the formulated SEE maximization problem is non-convex, it is first recast into a series of convex problems in order to obtain the optimal solution with a reasonable computational complexity.
Numerical results are used to verify the performance of the proposed algorithm and to reveal practical insights.
\end{abstract}

\section{Introduction}
Radio frequency energy harvesting (RF-EH)  technology has received much research  attention to provide energy for the low-power wireless nodes ubiquitously.
Moreover, the RF signals also enables the controllable and continuous energy transmission benefitting from its well-penetration property \cite{Dong2016, BiApri2015, ZengMay}.
As an important use case of RF-EH technology, the simultaneous wireless information and power transferring (SWIPT) systems were proposed to broadcast energy signals and deliver information signals simultaneously \cite{Varshneyjuly2008, ZhouNov.2013, ClerckxFeb.2018, HuangNov.}.
The concept of SWIPT systems was originally proposed by Varshney in \cite{Varshneyjuly2008}, where he studied the fundamental tradeoff within the \mbox{rate-energy} region in point-to-point (PtP) systems.
Then, the authors in \cite{ZhouNov.2013} proposed two practical receiver architectures, where the RF signals are split into two streams for energy harvesting module and information detecting module.
Moreover, the authors in \cite{ZhouNov.2013, ClerckxFeb.2018} investigated the \mbox{rate-energy} region of the PtP systems for linear energy harvester and non-linear energy harvester, respectively.
Since multiple antenna technology can enable the transmission of practical amount of energy, the research on SWIPT systems was extended to multiple-input-single-output (MISO) and multiple-input-multiple-output (MIMO) systems \cite{ShiFeb.2016, SonNov.2014, BoshkovskaMay2017}.

Though the broadcast character of wireless channels enables one-to-many energy delivery, it also increases the probability of legitimate information to be eavesdropped in the SWIPT systems \cite{LiuApr.2014, ShiMay2015, NgSept.2015}.
Therefore, the integration of physical layer security into the SWIPT systems becomes an emerging research topic.
For examples, the authors in \cite{LiuApr.2014} investigated the secrecy rate maximization (SRM) and harvested energy maximization (HEM) in the PtP systems with the multiple energy harvesting nodes (EHNs) co-located with eavesdroppers (EVEs).
Combining Charnes-Cooper transformation \cite{Charnes1962} with one-dimensional search method, the authors in \cite{LiuApr.2014} proposed optimal solutions to both SRM and HEM problems.
Then, the authors in \cite{ShiMay2015} studied the SRM in single-stream and multiple-stream MIMO PtP systems with perfect channel state information (CSI).
Using a semidefinite relaxation technique (SDR) and an inexact block coordinate descent method, the optimal and near-optimal solutions are obtained for the single-stream and multiple-stream cases.
The authors in \cite{NgSept.2015} studied the secure communication in the renewable energy powered distributed antenna systems with SWIPT capability, which allows each access point to exchange energy with the central processor.
By jointly designing the beamforming vector, artificial noise (AN) covariance matrix and energy exchange variables, the system power consumption is minimized in \cite{NgSept.2015}.

Due to the increased energy consumption, the wireless operators require new approaches to reduce their energy bills or improve the utilization efficiency of their purchased energy from the grid.
Hence, the energy efficiency optimization is one of the most important research issues for the future generation wireless communication systems \cite{WuAug.2017, LiuDec.2017}.
Although there are research efforts on energy efficiency optimization in SWIPT systems \cite{ShiFeb.2016, ShengApr.2016}, the secrecy energy efficiency (SEE) optimization in SWIPT systems has become an active research area recently \cite{MeiAug.2017}.
With perfect CSI, the authors in \cite{MeiAug.2017} investigated the SEE optimization problem in the MIMO PtP systems with multiple EVEs.
Using Dinkelbach method, they proposed an iterative algorithm to obtain the optimal SEE. However, their proposed algorithm cannot be applied to the SWIPT systems with multiple legitimated users (LUEs).

Since the perfect CSI is challenging to obtain in the MISO and MIMO systems, the authors in \cite{NgAug.2014} studied the system power minimization in the multiuser multiple-input-single-output-multiple-eavesdropper (MISOME)-SWIPT systems with imperfect CSI.
Specifically, they leveraged the energy signals and AN to secure the legitimate communication and satisfy the energy requirement.
Then, the authors extended the work \cite{NgAug.2014} and jointly considered the transmission power, energy harvesting efficiency and interference power leakage as a multiobjective optimization problem with imperfect CSI \cite{NgMay2016}.
In \cite{Lutobepublished}, the authors studied the impact of non-linear energy harvester on system power consumption in a multiuser \mbox{MISOME-SWIPT} system.
To our best knowledge, the robust SEE optimization problem in the multiuser MISOME-SWIPT systems with proportional secrecy rate constraints has not been reported in current literature.

In this work, we investigate the SEE optimization via joint design of beamforming vector and AN covariance matrix under imperfect CSI of EVEs and EHNs.
Moreover, we include the energy harvesting constraints and proportional secrecy rate constraints to guarantee the harvested energy at the EHNs and ensure the fairness among LUEs as in \cite{ShenNov.2005, BansalMar.2013}, respectively.
The formulated SEE optimization problem is more complicated than the traditional energy efficiency optimization problem \cite{ShiFeb.2016, ShengApr.2016} due to the non-convex proportional constraints and non-convex SEE function.
In order to develop tractable algorithms, we first simplify the SEE optimization problem by exploiting the structure of the underlying mathematical problem.
Based on the introduced parameter, we leverage the SDR technique coupled with a one-dimensional search method to address the simplified SEE optimization problem.
Numerical results are used to verify the performance of the proposed optimal algorithm.

\emph{Notations:}
Vectors and matrices are shown in bold lowercase letters and bold uppercase letters, respectively.
$\mathbb{C}^{N\times M}$ and $\mathbb{H}^{N}$ respectively denote all the $N\times M$ dimension complex value matrices and $N\times N$ Hermitian matrices.
$\left\|\cdot\right\|_{\F}$ and $\left|\cdot\right|$ are the Frobenius norm  and absolute value, respectively.
$\sim$ stands for ``distributed as''.
$\bm I_{N}$ and $\bm 0_{N\times M}$ denote, respectively, an $N$ dimensional identity matrix and a zero matrix with $N$ rows and $M$ columns.
The expectation of a random variable is denoted as $\mathds{E}\left[\cdot\right]$.
$\mbox{vec}\left[\bm W\right]$ obtains a vector by stacking the columns of $\bm W$ under the other.
$\left\{\bm{w}_n\right\}_{n \in \mathcal N}$ represents the set made of $\bm{w}_n$, $n \in \mathcal N$.
For a square matrix $\bm W$, $\bm W^{\H}$ and $\mbox{Tr}\left(\bm W\right)$ denote its conjugate transpose and trace, respectively.
$\bm W \succeq \bm 0$ and $\bm W \succ \bm 0$ respectively denote that $\bm W$ is a positive semidefinite and $\bm W$ is a positive definite matrix.

\section{System Model and Problem Formulation}
We consider downlink transmission of a \mbox{MISOME-SWIPT} system, which consists of one single BST, a set of LUEs, a set of EVEs and a set of EHNs.
Let $\mathcal N = \left\{1, 2, \ldots, N\right\}$, $\mathcal M = \left\{1, 2, \ldots, M\right\}$ and $\mathcal I = \left\{1, 2, \ldots, I\right\}$ denote the set of LUEs, set of EVEs and set of EHNs, respectively.
The BST is equipped with $N_t$ antennas for simultaneous information and energy transmission.
Each LUE is equipped with one single antenna to receive the legitimate information.
Each EVE, which is also equipped with one single antenna, can be roaming user from other wireless communication systems and searching for the services from the local BST.
Therefore, we leverage the AN to interrupt EVEs as well as guarantee energy harvesting requirement of EHNs.
Here, the EHNs are equipped with one single antenna and can be the passive sensors to monitor the status of the \mbox{MISOME-SWIPT} system.

At the BST, the information-bearing signal for the $n$-th LUE is given as $s_n$ with $\mathds{E}\[|s_n|^2\] = 1$.
Therefore, the transmission signal for the LUEs is denoted by $\sum\nolimits_{n = 1}^N {{\bm w_n}{s_n}}$, where $\bm w_n \in \mathbb{C}^{N_t\times 1}$ is the beamforming vector for the $n$-th LUE.
In order to secure the communication links between BST  and LUEs, the BST needs to transmit an AN vector $\bm q \in \mathbb{C}^{N_t\times 1}$ to reduce information leakage to the EVEs.
Hence, the transmission signal at the BST is given as $\bm x = \sum\nolimits_{n = 1}^N {{\bm w_n}{s_n}}  + \bm q$ where the AN vector $\bm q$ is modeled as a complex Gaussian random vector with mean zero and covariance matrix $\bm Q \in \mathbb{C}^{N_t\times N_t}$.
In particular, it is assumed that the information-bearing signals $\left\{s_n\right\}_{n \in \cal N}$ and the AN vector $\bm q$ are statistically independent.

The frame-based frequency non-selective fading channels with unit duration for each frame is considered.
Therefore, the words ``energy'' and ``power'' can be used interchangeably.
In each frame, the BST broadcasts information signals and jamming signals over the same spectrum band.
We assume that the MISOME-SWIPT system operates in time-division duplex mode.
Hence, the downlink transmission of BST can obtain the perfect CSI of LUEs via the uplink handshaking signals and channel reciprocity \cite{Dong2018a}.
Considering the Rayleigh fading channels, we have $\bm h_n \sim {\cal CN}\(\bm 0,  \Omega_{u, n}^{-1}\bm I_{N_t}\)$ where each entry of the vector $\bm h_n \in \mathbb{C}^{N_t\times 1}$ denotes channel coefficient for the $n$-th LUE.
Here, $\Omega_{u, n}$ is pathloss of the $n$-th LUE \cite{Goldsmith2005, DongApr.2016, Dong2018a}.
As a result, the received signals of $n$-th LUE is expressed as
\begin{equation}\label{eq:03}
y_{u, n} = \bm h_n^{\H}\bm w_ns_n +  \sumnk {\bm h_n^{\H}{\bm w_k}{s_k}} + \bm h_n^{\H}\bm q + z_{u, n}
\end{equation}
where $z_{u, n} \sim {\cal CN}\(0, \sigma^2_{u, n}\)$ is the additive white Gaussian noise (AWGN) with mean zero and variance $\sigma^2_{u, n}$ of the $n$-th LUE.

Based on the received signals in \eqref{eq:03}, the data rate of the $n$-th LUE is denoted as
\begin{equation}\label{eq:04}
R_{u, n} = {\cal BW}\log\( 1+ \mbox{SINR}_{u, n} \)
\end{equation}
where
\begin{equation}\label{eq:05}
\mbox{SINR}_{u, n} = \frac{\Tr\(\bm H_n\bm W_n\)}{\sum\limits_{k=1, k \neq n}^N \Tr\(\bm H_n \bm W_k\)  + \Tr\(\bm H_n\bm Q\) + {\sigma _{u, n}^2}}
\end{equation}
with $\bm H_n \triangleq \bm h_n\bm h_n^{\H}$ and $\bm W_n \triangleq \bm w_n\bm w_n^{\H}$.
In addition, the rank of each beamforming matrix $\bm W_n$ is upper bounded as $\Rank\(\bm W_n\) \le 1$.
The term ${\cal BW}$ denotes the bandwidth of the system.

Besides, the received signal at the $m$-th EVE is denoted as
\begin{equation}\label{eq:06}
y_{e, m} = {\bm g_{e, m}^{\H}}{\bm w_n}{s_n} + \sumnk {{\bm g_{e, m}^{\H}}{\bm w_k}{s_k}}  + {\bm g_{e, m}^{\H}}\bm q + z_{e, m}
\end{equation}
where $z_{e, m} \sim {\cal CN}\(0, \sigma^2_{e, m}\)$ is the AWGN with mean zero and variance $\sigma^2_{e, m}$ at the $m$-th EVE, and $\bm g_{e,m} \in \mathbb{C}^{N_t \times 1}$ is the channel coefficient vector of the $m$-th EVE.

The information leakage to the $m$-th EVE of the $n$-th LUE by treating the interference as noise is denoted as \cite{Goldsmith2005}
\begin{equation}\label{eq:07}
R_{e, m \rightarrow n} = {\cal BW}\log\( 1 + \SINR_{e, m \rightarrow n}\)
\end{equation}
where
\begin{equation}\label{eq:08}
\SINR_{e, m\rightarrow n} = \frac{\Tr\(\bm G_{e, m}\bm W_n\)}{\sum\limits_{k=1, k \neq n}^N \Tr\(\bm G_{e, m} \bm W_k\)  + \Tr\(\bm G_{e, m}\bm Q\) + {\sigma _{e, m}^2}}
\end{equation}
and where $\bm G_{e, m} \triangleq \bm g_{e, m}\bm g_{e, m}^{\H}$.

Denote the channel vector of the $i$-th EHN as $\bm g_{h, i} \in \mathbb{C}^{N_t\times 1}$.
The received baseband signal of the $i$-th EHN is given as
\begin{equation}\label{eq:09}
y_{h, i} = \sumn\bm g_{h, i}^{\H}\bm w_n s_n + \bm g_{h, i}^{\H}\bm q + z_{h, i}
\end{equation}
where $z_{h, i} \sim {\cal CN}\(0, \sigma_{h,i}^2\)$ is the AWGN at the $i$-th EHN.

Neglecting the harvested energy from the AWGN, the amount of harvested energy of the $i$-th EHN is denoted as
\begin{equation}\label{eq:10}
P_{h, i} = \xi_{h, i}\(\sumn\Tr\(\bm G_{h, i}\bm W_n\) + \Tr\(\bm G_{h, i}\bm Q\)\)
\end{equation}
where $\bm G_{h, i} \triangleq \bm g_{h, i}\bm g_{h, i}^{\H}$, and $\xi_{h, i} \in \(0, 1\]$ is the energy conversion efficiency.

In this work, we consider a conservative secrecy rate where the received SINR of the $m$-th EVE to detect the $n$-th LUE's information is confined to be smaller than or equal to a predefinied threshold $R_{e, m \rightarrow n}^{\REQ}$.
The secrecy rate of the $n$-th LUE is denoted as \cite{LiuApr.2014, ShiMay2015}
\begin{equation}\label{eq:11}
R_{u, n}^{\SEC} = \( R_{u, n} - \max\limits_{m \in \cal M}R^{\REQ}_{e, m\rightarrow n}\)^+
\end{equation}
with
\begin{equation}\label{eq:12}
R_{e, m\rightarrow n} \le R^{\REQ}_{e, m\rightarrow n}
\end{equation}
where $\(\cdot\)^+ = \max\(\cdot, 0\)$.

Furthermore, the total power consumption of the MISOME-SWIPT system is comprised of the power of downlink beamforming, power of AN and power of circuit.
Therefore, the expression of the power consumption is given as
\begin{equation}\label{eq:13}
P^{\TOT} = \frac{1}{\phi }\left( {\sum\limits_{n = 1}^N \Tr\(\bm W_n\)  + \Tr\left( \bm Q \right)} \right) + {P^{\C}}
\end{equation}
where the constant circuit power consumption $P^{\C}$ is given as $P^{\C} = P^{\mbox{\tiny SP}}\(0.87 + 0.1N_t + 0.03N_t^2\)$ since the circuit power consumption is positively correlated to the number of antennas \cite{AttarOct.2011}.
Here,  the term $P^{\mbox{\tiny SP}}$ denotes the baseband processing power consumption.

In this work, the secrecy energy efficiency is defined as the ratio of the sum secrecy rates of LUEs over the total power consumption of BST .
Based on \eqref{eq:11} and \eqref{eq:13}, the SEE of \mbox{MISOME-SWIPT} system is given as
\begin{equation}\label{eq:14}
\mbox{EE}\left(\left\{\bm W_n\right\}_{n \in \cal N}, \bm Q\right) = \frac{\sumn R_{u, n}^{\SEC}}{{P^{\TOT}}}
\end{equation}
whose unit is joules/nat.
Different from \cite{ShiFeb.2016}, we ignore the harvested energy in the system total power consumption and investigate the SEE maximization.
This is due to the fact that the harvested energy is usually stored in the integrated battery of EHNs and cannot reduce the energy expenditure of the wireless operators.

Since the EVEs and EHNs do not send the handshaking signals to the BST frequently, the CSI for EVE-BST links and EHN-BST links are challenging  to estimate.
Motivated by \cite{WangAug.2009, SunAug.2016}, we use the bounded CSI error model to formulate the uncertainty of channel vectors $\bm g_{e, m}$ and $\bm g_{h, i}$ as
\begin{align}
\bm g_{e, m} &= \bar{\bm g}_{e, m} + \Delta\bm g_{e, m}, \left\|\Delta\bm g_{e, m} \right\|_{\F} \le \Theta_{e, m}, \forall m \\
\bm g_{h, i} &= \bar{\bm g}_{h, i} + \Delta{\bm g}_{h, i}, \left\|\Delta{\bm g}_{h, i} \right\|_{\F} \le \Theta_{h, i}, \forall i
\end{align}
where $\bar{\bm g}_{e, m} \sim (\bm 0, \Omega_{e,m}^{-1}\bm I_{N_t})$ and $\Delta\bm g_{e, m} \in \mathbb{C}^{N_t \times 1}$ are, respectively, the estimated channel coefficient vector and channel uncertainty of the $m$-th EVE.
Similarly, $\bar{\bm g}_{h, i} \sim (\bm 0, \Omega_{h,i}^{-1}\bm I_{N_t})$ and $\Delta{\bm g}_{h, i}$ are the estimated channel coefficient vector and channel uncertainty vector of the $i$-th EHN.
The positive constants $\Theta_{e, m}$ and $\Theta_{h, i}$ denote the radii of uncertainty regions of $\bm g_{e, m}$ and $\bm g_{h, i}$, respectively.
Here, the channel uncertainty vectors $\Delta\bm g_{e, m}$ and $\Delta{\bm g}_{h, i}$ capture the joint effect of estimation errors and time-varying characteristics of wireless channels.

Our objective is to minimize SEE via joint design of beamforming vector and AN covariance matrix in the MISOME-SWIPT system in a centralized way at the BST.
The SEE maximization problem is formulated as
\begin{subequations}\label{eq:15}
\begin{align}
\hspace{-0.6 cm}
\max\limits_{\left\{ \bm W_n \right\}_{n \in \cal N},\bm Q} &\; \mbox{EE}\( \left\{ {{\bm W_n}} \right\}_{n \in \cal N},\bm Q \) \label{eq:15a}\\
\mbox{s.t.}
&\; \max\limits_{\left\|\Delta\bm g_{e, m}\right\|_{\F} \le \Theta_{e, m}} R_{e, m\rightarrow n} \le R_{e, m\rightarrow n}^{\REQ}, \forall m, n \label{eq:15d}\\
&\; \min\limits_{\left\|\Delta\bm G_{h, i}\right\|_{\F} \le \Theta_{h, i}} P_{h, i} \ge P_{h, i}^{\REQ}, \forall i \label{eq:15e}\\
&\; {R_{u, 1}^{\SEC}}: \ldots :{R_{u, N}^{\SEC}} = {\varphi _1}: \ldots :{\varphi _N}\label{eq:15c}\\
&\; \sum\limits_{n = 1}^N \Tr\(\bm W_n\) +\Tr\left(\bm Q\right) \le {P^{\max }}\label{eq:15b}\\
&\; \bm Q \succeq \bm 0, \bm W_n \succeq \bm 0,  \forall n \label{eq:15g} \\
&\; \Rank\(\bm W_n\) \le 1, \forall n \label{eq:15f}
\end{align}
\end{subequations}
where the constant $P^{\max}$ in the constraints in \eqref{eq:15b} is the power budget of the BST due to the circuit limitation; the constraints in \eqref{eq:15c} is used to guarantee the proportional fairness on the secrecy rate of LUEs with $\sum\nolimits_{n=1}^N \varphi_n = 1$ and $\varphi_n \ge 0$ \cite{ShenNov.2005, BansalMar.2013};
the information-leakage constraints and the harvested-power constraints are respectively in \eqref{eq:15d} and \eqref{eq:15e}.
Here, the term $P_{h, i}^{\REQ}$ in \eqref{eq:15e} denotes the energy requirement of the $i$-th EHN.

\begin{remark}
Although we use the linear energy harvester, it is easy to justify that, by properly setting the threshold $P_{h, i}^{\REQ}$, the non-linear energy harvester model can be incorporated in \eqref{eq:15d} \cite{BoshkovskaDec.2015, DongApr.2016}. Based on the model in \cite{BoshkovskaDec.2015}, harvesting $P_{h, i}^{\REQ}$ power requires the input power
\begin{equation*}
\hat P_{h, i}^{\REQ} = \frac{\xi_{h, i}}{a_{h, i}}\log\left(\frac{M_{h, i} + P_{h, i}^{\REQ}\exp\left(a_{h, i}b_{h, i}\right)}{M_{h, i} - P_{h, i}^{\REQ}}\right)
\end{equation*}
where $M_{h, i}$, $a_{h, i}$ and $b_{h, i}$ are the shaping parameters \cite{BoshkovskaDec.2015}.
\end{remark}

\section{Optimal SEE Maximization}

In order to deal with the infinite amount of information-leakage constraints in \eqref{eq:15d} and energy-harvesting constraints in \eqref{eq:15e}, we first review the $\cal S$-procedure \cite{boyd2004convex} in Lemma \ref{le:01}.

\begin{lemma}[$\cal S$-procedure \cite{boyd2004convex}]\label{le:01}
Denote the functions $f_1\(\bm x\)$ and $f_2\(\bm x\)$ as $f_1\(\bm x\) = \bm x^{\H}\bm A_1\bm x + \bm b_1^{\H}\bm x + \bm x^{\H}\bm b_1 + c_1$ and $f_2\(\bm x\) = \bm x^{\H}\bm A_2\bm x + \bm b_2^{\H}\bm x + \bm x^{\H}\bm b_2 + c_2$ with $\bm A_1 = \bm A_1^{\H}$ and $\bm A_2 = \bm A_2^{\H}$.
Then, the condition $f_1\(\bm x\) \le 0 \Rightarrow f_2\(\bm x\) \le 0$ holds if and only if there is $\eta \ge 0$ such that
\begin{equation}\label{eq:le1:01}
\eta\[ {\begin{array}{*{20}{c}}
{\bm A_1}&{\bm b_1}\\
{\bm b_1^{\H}}&{c_1}
\end{array}} \] -
\[ {\begin{array}{*{20}{c}}
{\bm A_2}&{\bm b_2}\\
{\bm b_2^{\H}}&{c_2}
\end{array}} \] \succeq \bm 0
\end{equation}
when there exists a point $\bm x$ such that $f_k\(\bm x\) < 0$, $k = 1, 2$.
\end{lemma}

Based on \eqref{eq:07} and \eqref{eq:08}, we obtain an equivalent form of the information leakage constraints in \eqref{eq:12} as
\begin{equation}\label{eq:16}
\bm g_{e, m}^{\H} \bm X_n \bm g_{e, m} \le \sigma_{e, m}^2, \forall \Delta\bm g_{e, m}, \forall m, n
\end{equation}
where $\bm X_n \triangleq \frac{1}{1 - \exp\(-\widetilde{R}^{\REQ}_{e, m\rightarrow n}\)}\bm W_n - \sum\nolimits_{k=1}^N \bm W_k - \bm Q$ with $\widetilde{R}^{\REQ}_{e, m\rightarrow n} =  \frac{R_{e, m\rightarrow n}^{\REQ}}{\cal BW}$ and $\left\|\Delta \bm g_{e, m}\right\|_{\F}^2 \le \Theta^2_{e, m}$.

Applying Lemma \ref{le:01} to \eqref{eq:16}, we obtain an equivalent form of information-leakage constraints in \eqref{eq:15d} as
\begin{equation}\label{eq:17}
\[ {\begin{array}{*{20}{c}}
\zeta_{m, n}\bm I_{N_t}& \bm 0 \\
\bm 0 & \sigma_{e,m}^2 - \zeta_{m, n}\Theta_{e,m}^2
\end{array}} \] - \widetilde{\bm G}_{e, m}^{\H} \bm X_n \widetilde{\bm G}_{e, m} \succeq \bm 0, \forall m, n
\end{equation}
where $\widetilde{\bm G}_{e, m} \triangleq \[\bm I_{N_t}, \bar{\bm g}_{e,m}\]$ and $\zeta_{m, n} \ge 0$ is the introduced auxiliary variable, $m \in \cal M$ and $n \in \cal N$.

Similarly, the equivalent form of \eqref{eq:15e} is denoted as
\begin{equation}\label{eq:21}
\bm g^{\H}_{h, i}\bm Y\bm g_{h, i} \ge \frac{P_{h, i}^{\REQ}}{\xi_{h, i}}, \forall \Delta \bm g_{h, i}, \forall i
\end{equation}
with $\Delta\bm g^{\H}_{h, i}\Delta\bm g_{h, i} \le \Theta^2_{h, i}$, $i \in \cal I$.
Here, $\bm Y \triangleq \sum\nolimits_{n=1}^N \bm W_n + \bm Q$.

Then, applying Lemma \ref{le:01} to \eqref{eq:21} and performing some algebraic manipulations, we obtain an equivalent form of energy-harvesting constraints in \eqref{eq:15e} as
\begin{equation}\label{eq:22}
\[ {\begin{array}{*{20}{c}}
\eta_{i}\bm I_{N_t}& \bm 0 \\
\bm 0 & -\frac{P^{\REQ}_{h, i}}{\xi_{h, i}} - \eta_{i}\Theta_{h,i}^2
\end{array}} \] +
\widetilde{\bm G}_{h,i}^{\H}\bm Y\widetilde{\bm G}_{h,i} \succeq \bm 0, \forall i
\end{equation}
where $\widetilde{\bm G}_{h,i} = \[\bm I_{N_t}, \bm g_{h,i}\]$ and $\eta_i \ge 0$ is the introduced auxiliary variable, $i \in \cal I$.

Next, to handle the non-convex proportional secrecy rate constraints in \eqref{eq:15c}, we introduce an auxiliary constant $t$ such that ${\theta _n}\left( t \right) = \exp \(\frac{\varphi_n t}{\cal BW} + \max\nolimits_{m \in \cal M} \widetilde{R}^{\REQ}_{e, m\rightarrow n}\) - 1$.
Then, we relax the \mbox{non-convex} proportional secrecy rate constraints in \eqref{eq:15c} into a set of convex constraints as
\begin{multline}\label{eq:23}
\frac{1 + \theta_n\(t\)}{\theta_n\(t\)}\Tr\(\bm H_n \bm W_n\) \\
\ge \sumk \Tr\(\bm H_n \bm W_k\) + \Tr\(\bm H_n \bm Q\) + \sigma_{u, n}^2, \forall n.
\end{multline}

Since the constraints in \eqref{eq:23} are obtained via convex relaxation, we are motivated to verify the tightness of relaxation in Proposition \ref{pr:01}.

\begin{proposition}\label{pr:01}
Suppose that the constraints in \eqref{eq:23} are inactive under the optimal $\{\bm W_n^*\}_{n \in \cal N}$ and $\bm Q^*$ to \eqref{eq:15}, another optimal solution can be constructed as
\begin{subequations}
\begin{align}
\widehat{\bm W}_n^* &= \bm W_n^* - \Delta \bm W_n^*, \forall n \label{eq:18a}\\
\widehat{\bm Q}^* &= \bm Q^* + \sumn \Delta \bm W_n^*  \label{eq:18b}
\end{align}
\end{subequations}
with constraints in \eqref{eq:23} being active.
\end{proposition}
\begin{IEEEproof}
Due to the space limitation, the detailed proof will be provided in the extended version of this conference paper \cite{dong2018jsac}.
\end{IEEEproof}

Proposition \ref{pr:01} indicates that the relaxation in \eqref{eq:23} is tight, and the constraints in \eqref{eq:23} are equivalent to the constraints in \eqref{eq:15c}.
Therefore, we are motivated to use the constraints in \eqref{eq:23} to replace non-convex constraints in \eqref{eq:15c} to seek low complexity algorithm for the problem in \eqref{eq:15}.

Based on the previous manipulations, we can now convexify the feasible region of the problem \eqref{eq:15} via SDR technique.
By dropping the rank constraints in \eqref{eq:15f}, the feasible region of problem \eqref{eq:15} is transformed into a convex one as
\begin{subequations}\label{eq:24}
\begin{align}
\mathop {\min }\limits_{\scriptstyle{\left\{ {\bm W_n} \right\}_{n \in {\cal N}}}, \bm Q, \{\eta_i\}_{i \in \cal I}\hfill\atop
\scriptstyle{\left\{ {{\zeta _{m,n}}} \right\}_{m \in {\cal M},n \in {\cal N}}, t}\hfill} &\;
\frac{1}{t}P^{\TOT}\(\left\{\bm W_n\right\}_{n \in \cal N}, \bm Q\) \label{eq:24a}\\
\mbox{s.t.}&\; \zeta_{m, n} \ge 0, \eta_i \ge 0, \forall m, n, i \label{eq:24b}\\
&\; \eqref{eq:15b}, \eqref{eq:15g}, \eqref{eq:17}, \eqref{eq:22} \mbox{ and } \eqref{eq:23}. \label{eq:24d}
\end{align}
\end{subequations}

We have two observations of the problem \eqref{eq:24}: 1) the power constraint in \eqref{eq:15b} and objective function in \eqref{eq:24a} have a common term $\sum\nolimits_{n=1}^N\Tr\(\bm W_n\) + \Tr\(\bm Q\)$; and 2) the feasible region of the problem \eqref{eq:24} is a convex hull of the feasible region of the problem \eqref{eq:15}.

Based on the first observation, we are motivated to remove the power constraint in \eqref{eq:15b} to simplify the optimization problem.
Dropping the power constraint in \eqref{eq:15b}, we solve the problem \eqref{eq:24} for a value of $t$ as
\begin{subequations}\label{eq:26}
\begin{align}
\mathop {\min }\limits_{\scriptstyle{\{\bm W_n\}_{n \in {\cal N}}}, \bm Q, \{\eta_i\}_{i \in \cal I}\hfill\atop
\scriptstyle{\left\{ {{\zeta _{m,n}}} \right\}_{m \in {\cal M},n \in {\cal N}}}\hfill} &\;
\sum\limits_{n=1}^N\Tr\(\bm W_n\) + \Tr\(\bm Q\) \label{eq:26a}\\
\mbox{s.t.}&\; \zeta_{m, n} \ge 0, \forall m, n \label{eq:26b}\\
&\; \eta_i \ge 0, \forall i \label{eq:26c} \\
&\; \eqref{eq:15g}, \eqref{eq:17}, \eqref{eq:22} \mbox{ and } \eqref{eq:23}. \label{eq:26d}
\end{align}
\end{subequations}

Since the problem \eqref{eq:24} is feasible if and only if the optimal value of \eqref{eq:26} is feasible and the optimal value of \eqref{eq:26a} is less than or equal to $P^{\max}$.
Hereinafter, without loss of generality, we assume that the problem \eqref{eq:26} is feasible.
In the problem \eqref{eq:26}, the power constraints in \eqref{eq:15b} is a function of $t$.
We will leverage the following Proposition \ref{pr:02} to justify that the optimal value of \eqref{eq:24} can be obtained via the proposed two-stage algorithm.

\begin{proposition}\label{pr:02}
Let $f\(t\) = \sum\nolimits_{n=1}^N\Tr\(\bm W_n^*\(t\)\) + \Tr\(\bm Q^*\(t\)\)$ denote the optimal power consumption of \eqref{eq:26a} given $t$ where $\left\{\bm W_n^*\(t\)\right\}_{n \in \cal N}$ and $\bm Q^*\(t\)$ are, respectively, the optimal beamforming matrices and AN covariance matrix to \eqref{eq:26} given $t$.
The optimal power consumption of \eqref{eq:26a} increases monotonically with respect to the value of $t$.
\end{proposition}
\begin{IEEEproof}
Due to the space limitation, the detailed proof will be provided in the extended version of this conference paper \cite{dong2018jsac}.
\end{IEEEproof}

With the Proposition \ref{pr:02}, we can claim that there exists only one value of $t^{\max}$ such that the optimal power consumption $f\(t^{\max}\) = P^{\max}$.
The value of $t^{\max}$ can be obtained  via a one-dimensional search.

Now, we investigate the optimality of SDR technique.
Based on the second observation, we conclude that the remaining challenge to solve the problem is to deal with the non-convex objective function \eqref{eq:24a} with respect to the value of $t$, beamforming matrices $\{\bm W_n\}_{n \in \cal N}$ and AN covariance matrix $\bm Q$.
Note that the objective function \eqref{eq:24a} is convex with respect to the beamforming matrices $\{\bm W_n\}_{n \in \cal N}$ and AN covariance matrix $\bm Q$.
Therefore, we propose a two-stage algorithm to solve the problem in \eqref{eq:24}: 1) solving the problem \eqref{eq:24} with given $t$; 2) performing a one-dimensional search to obtain the optimal $t^*$.
Given the optimal value of $t^*$, the optimal value of \eqref{eq:24} serves as a lower bound of the optimal value of \eqref{eq:15} due to the SDR operation.
Thus, we are motivated to study the gap between the optimal values of \eqref{eq:24} and \eqref{eq:15} in Proposition \ref{pr:03}.

\begin{proposition}\label{pr:03}
Given a value of $t$, there exists an optimal solution $\(\{\bm W_n^*\}_{n\in \cal N}, \bm Q^*, \{\zeta_{m, n}^*\}_{m \in {\cal M}, n \in {\cal N}}, \{\eta_i^*\}_{i \in \cal I}\)$ to the problem \eqref{eq:26} with the rank of beamforming matrix $\bm W_n^*$ satisfying
\begin{equation}\label{eq:pr:01}
\Rank\(\bm W_n^*\) \le 1, \forall n.
\end{equation}
\end{proposition}
\begin{IEEEproof}
Due to the space limitation, the detailed proof will be provided in the extended version of this conference paper \cite{dong2018jsac}.
\end{IEEEproof}

Based on Proposition \ref{pr:03}, we conclude that the optimal value of \eqref{eq:24a} equals to that of \eqref{eq:15a} given the optimal value of $t^*$.
Moreover, the optimal value of $t^*$ can be obtained via a one-dimensional search based on the monotonicity of objective function \eqref{eq:24a} with respect to $t$.
Hence, we propose \textbf{Algorithm} \ref{alg:01} that is based on the SDP empowered two-stage beamforming and artificial jamming (SDP-TsBAJ) algorithm.

\begin{algorithm}[ht]\small
  \centering
  \caption{SDP Empowered Two-Stage Beamforming and Artificial Jamming Algorithm}\label{alg:01}
  \begin{algorithmic}[1]
  \State BST sets the iteration index $\tau = 0$ and the stop threshold $\epsilon$.
  \State BST initializes $t = 0$ and $\Delta t$
  \Repeat
  \State BST obtains ${\cal OBJ}^{\OPT}\(t\)$ via solving \eqref{eq:26}
  \State $t \leftarrow t + \Delta t$
  \Until{$\sum\nolimits_{n=1}^N\Tr\(\W_n\) + \Tr\(\Q\) > P^{\max}$}
  \State BST obtains the optimal value of \eqref{eq:15} and the beamforming matrices $\left\{\bm W_n^*\right\}_{n \in \cal N}$ and AN matrix $\bm Q^*$
  \If{$\Rank\(\bm W_n^*\) > 1$}
  \State BST performs the rank-one recovery procedure in \eqref{eq:18a} and \eqref{eq:18b}
  \EndIf
  \end{algorithmic}
\end{algorithm}

\section{Numerical Results}
In this section, we present simulation results to demonstrate the performances of the proposed algorithms.~The pathloss model follows indoor channels as $\Omega = 17.3 + 24.9\log_{10}f_c + 38.3\log_{10} d$ dB, where $f_c$ and $d$ are respectively the carrier frequency and link distance \cite{3gpp_pathloss}.
The values of $\Theta^2_{e, m}$ and $\Theta^2_{h, i}$ are set as $0.05\Omega_{e,m}$ and $0.05\Omega_{h,i}$, respectively.
The simulation parameters are set in Table \ref{table:03}.

\begin{table}[htbp]\scriptsize
  \caption{Simulation Parameters Setting}\label{table:03}
  \centering
  \begin{tabular}{|l|c|}
    \hline
    {\textbf{Parameters}} & \textbf{Values} \\\hline
    {Carrier frequency and system bandwidth},  $f_c$, ${\cal BW}$ & 900 MHz, 200 KHz  \\\hline
    {Number of LUEs, EVEs and EHNs},  $N$, $M$, $I$ & 3, 2, 2 \\\hline
    {Number of antennas at BST},  $N_t$ & 7 \\\hline
    {Power of AWGN},  $\sigma_{n, u}^2$, $\sigma_{e, m}^2$, $\sigma_{h, i}^2$ & -30 dBm \\\hline
    {Maximum Tx power of BS},  $P^{\max}$ & 43 dBm \\\hline
    {Required harvested power of the $i$-th EHN},  $P_{h, i}^{\REQ}$ & -5 dBm \\\hline
    {Energy conversion efficiency of the $i$-th EHN},  $\xi_{h, i}$ & 0.8 \\\hline
    {Rate of auxiliary information of the $n$-th LUE},   $R^{\REQ}_{e, n}$ & 100 Knats/sec \\\hline
    {Secrecy rate ratios},  $\varphi_1, \varphi_2, \varphi_3$ & 0.4, 0.3, 0.3\\\hline
    {Amplifier efficiency}, $\phi$ & 0.8 \\\hline
    {Distances of EVEs and EHNs to BST},  $d_{e, m}$ and $d_{h, i}$ & 8 m and 6 m \\\hline
    {Distances of LUEs, EVEs and EHNs to BST},  {$d_{u, n}$} & {16 m, 19 m and 22 m} \\\hline
  \end{tabular}%
\end{table}%

In order to demonstrate the effectiveness of the proposed algorithms, we consider another set of baseline schemes in the remaining simulations, where the BST solves the SRM problem with PSR constraints.
Hereinafter, we name the considered  baselines as SRM-SDP, SRM-ZFBF and SRM-MRT-ZFBF algorithms.

\begin{figure}[htb]
\vspace{-0.2 cm}
\centering
  \includegraphics[width= 2.6 in]{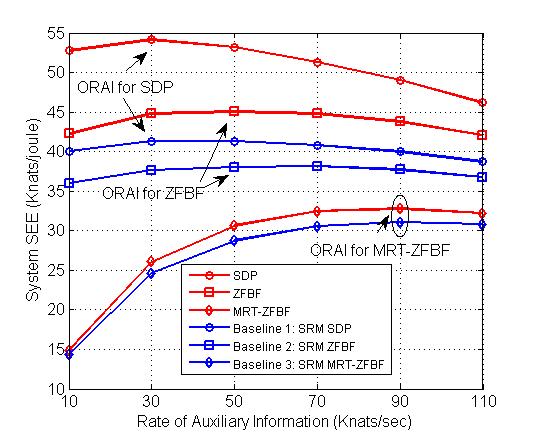}
  \caption{System SEE v.s. rate of auxiliary information.}\label{fg:sm:06a}
\vspace{-0.5 cm}
\end{figure}

Figure \ref{fg:sm:06a} shows that the system SEE varies with the rate of auxiliary information.
Here, the relative gap is obtained as
\begin{equation}
\mbox{Relative Gap} = \frac{\mbox{SEE}_{\rm SDP-TsBAJ} - \mbox{SEE}_{\pi}}{\mbox{SEE}_{\rm SDP-TsBAJ}}
\end{equation}
where $\pi$ can be ZFBF-TsBAJ, MRT-ZFBF-TsBAJ, SRM-SDP, SRM-ZFBF and SRM-MRT-ZFBF algorithms.

We observe that the system SEE increases monotonically with the rate of auxiliary information.
After reaching the peak, the system SEE starts to decreases with the rate of auxiliary information.
This observation can be explained as follows: 1) a smaller rate of auxiliary information ($\le 30$ Knats/sec for SDP-TsBAJ, $\le 50$ Knats/sec for ZFBF-TsBAJ and $\le 90$ Knats/sec for MRT-ZFBF-TsBAJ) restricts the flexibility in enhancing the sum secrecy rate; and 2) larger rate of auxiliary information ($> 30$ Knats/sec for SDP-TsBAJ, $> 50$ Knats/sec for ZFBF-TsBAJ and $> 90$ Knats/sec for MRT-ZFBF-TsBAJ) sacrifices the rate of main information.
Therefore, we conclude that there exists an optimal rate of auxiliary information (ORAI).
For example, the ORAI for SDP based algorithms (SDP-TsBAJ and SRM-SDP algorithms) is around $30$ Knats/sec as shown in Fig. \ref{fg:sm:06a}.
The ORAIs for the ZFBF based algorithms (ZFBF-TsBAJ and SRM-ZFBF algorithms) and MRT-ZFBF based algorithms (MRT-ZFBF-TsBAJ and SRM-MRT-ZFBF algorithms) are, respectively, $50$ Knats/sec and $90$ Knats/sec as shown in Fig. \ref{fg:sm:06a}.

\begin{figure}[htb]
\vspace{-0.5 cm}
\centering
  \includegraphics[width= 2.6 in]{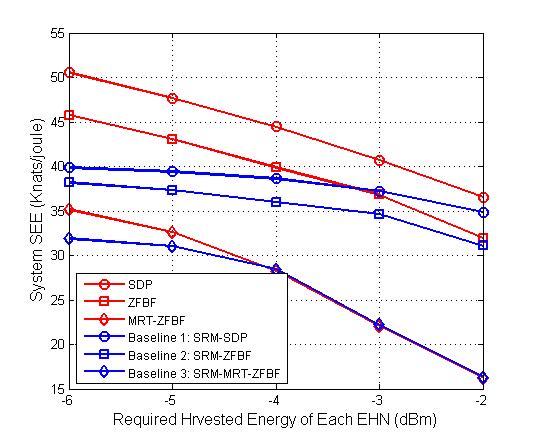}
  \vspace{-0.1 cm}
  \caption{The system SEE v.s. requirement of harvested energy.}\label{fg:sm:03a}
  \vspace{-0.5 cm}
\end{figure}

Figure \ref{fg:sm:03a} shows the variation of system SEE with minimum harvested energy.
We observe that the system SEE decreases with minimum harvested energy for the SDP-TsBAJ, ZFBF-TsBAJ algorithm and MRT-ZFBF-TsBAJ algorithms as well as three baseline algorithms.
This is a result of two facts: 1) a higher RF-EH demand leads to a higher power consumption at BST; and 2) the system secrecy rate becomes smaller with the higher RF-EH demand.
From Fig. \ref{fg:sm:03a}, we also observe that performance gap between SDP-TsBAJ and SRM-SDP diminishes with the increase in minimum harvested energy.
This is due to the fact that a higher RF-EH demand consumes more power on the energy delivery to EHNs such that the probability of BST draining its power budget becomes higher.
Therefore, the SEE maximization problem is reduced  to the SRM maximization problem when the minimum harvested energy increases.

\section{Conclusion}
We developed the SDP-TsBAJ algorithm for the SEE maximization problem via the joint design of beamforming vector and AN covariance matrix in the MISOME-SWIPT systems.
We included the proportional secrecy rate constraints to guarantee the fairness among the LUEs in any time scale.
Due to the non-convexity in the proportional secrecy rate constraints and SEE function, we exploited the structure of the formulated problem to propose efficient SDP-TsBAJ algorithm.
Numerical results show that the performance gap between the SDP-TsBAJ algorithm and MRT-ZFBF-TsBAJ algorithm decreases as the increasing information leakage rate.
Thus, there exists a tradeoff between the computational complexity and the SEE performance for the two algorithms.

\balance
\bibliographystyle{IEEEtran}
\bibliography{dyj_bib}
\end{document}